# Neu-RadBERT for Enhanced Diagnosis of Brain Injuries and Conditions

*SINGH, Manpreet[1]

*MACRAE, Sean[2]

WILLIAMS, Pierre-Marc[2]

HUNG, Nicole[2]

ARAUJO DE FRANCA, Sabrina[1]

LETOURNEAU-GUILLON, Laurent[2,3] orcid : 0000-0002-6325-7538

CARRIER, François-Martin[2,4]

LIU, Bang[5]

CAVAYAS, Yiorgos Alexandros[1,2,6#]

(1) Équipe de Recherche en Soins Intensifs, Centre de recherche du Centre intégré universitaire de santé et de services sociaux du Nord-de-l'Île-de-Montréal.
(2) Faculté de Médecine, Université de Montréal
(3) Department of Radiology, Centre Hospitalier de l'Université de Montréal
(4) Department of Anesthesia, Centre Hospitalier de l'Université de Montréal
(5) Applied Research in Computer Linguistics Laboratory, Department of Computer Science and Operations Research, Université de Montréal
(6) Division of Critical Care Medicine, Department of Medicine, Hôpital du Sacré-Coeur de Montréal

*Both have contributed equally and should be considered as first authors

#Corresponding author

Dr Yiorgos Alexandros Cavayas

Division of Critical Care Medicine, Department of Medicine,

Hôpital du Sacré-Coeur de Montréal,

5400 Boulevard Gouin Ouest,

Montreal, Quebec

H4J 1C5 (Canada)

yiorgos.alexandros.cavayas@umontreal.ca

Funding:

Dr Cavayas, Dr Letourneau-Guillon and Dr Carrier are supported by the Fonds de recherche du Québec – Santé. This project was funded by the Canadian Institutes of Health Research.

Conflicts of interest:

None of the authors had any conflicts of interests to declare.

# ABSTRACT

**Objective**: We sought to develop a classification algorithm to extract diagnoses from free-text radiology reports of brain imaging performed in patients with acute respiratory failure (ARF) undergoing invasive mechanical ventilation.

**Methods**: We developed and fine-tuned Neu-RadBERT, a BERT-based model, to classify unstructured radiology reports. We extracted all the brain imaging reports (computed tomography and magnetic resonance imaging) from MIMIC-IV database, performed in patients with ARF. Initial manual labelling was performed on a subset of reports for various brain abnormalities, followed by fine-tuning Neu-RadBERT using three strategies: 1) baseline RadBERT, 2) Neu-RadBERT with Masked Language Modeling (MLM) pretraining, and 3) Neu-RadBERT with MLM pretraining and oversampling to address data skewness. We compared the performance of this model to Llama-2-13B, an autoregressive LLM.

**Results**: The Neu-RadBERT model, particularly with oversampling, demonstrated significant improvements in diagnostic accuracy compared to baseline RadBERT for brain abnormalities, achieving up to 98.0% accuracy for acute brain injuries. Llama-2-13B exhibited relatively lower performance, peaking at 67.5% binary classification accuracy. This result highlights potential limitations of current autoregressive LLMs for this specific classification task, though it remains possible that larger models or further fine-tuning could improve performance.

**Conclusion**: Neu-RadBERT, enhanced through target domain pretraining and oversampling techniques, offered a robust tool for accurate and reliable diagnosis of neurological conditions from radiology reports. This study underscores the potential of transformer-based NLP models in automatically extracting diagnoses from free text reports with potential applications to both research and patient care.

**Introduction**

In the fields of radiology and clinical medicine, unstructured reports carry tremendous potential to enhance both patient care and medical research. A significant portion of healthcare information, from clinical notes to patient-reported outcomes, resides in free text form, which traditional data analysis tools cannot easily interpret. Such information has major potential for clinical and research applications ranging from new discoveries on previously hidden associations to the development of various models. However, manually extracting information from hundreds if not thousands of unstructured reports is often impractical due to the time-consuming and resource-intensive nature of this task. Automated extraction of information into a well-defined construct is required to enhance the potential of those reports.

Previous approaches to automate this process have included complex rule-based systems. These systems typically employ text parsing with regular expressions and strategies to classify negation, such as identifying pertinent negatives that may be classified as positive findings. Specific advancements in the field of natural language processing (NLP) have revolutionized the use of unstructured patient care data. These include the use of transformer-based models like Bidirectional Encoder Representations from Transformers (BERT) (Devlin et al. 2018) but also, more recently, other large language models (LLMs) such as the GPT-n series, PaLM, and Llama (Siu 2023). While the masked token method of BERT-based models (Y. Liu et al. 2023) is classically better suited for text classification, these new LLMs have demonstrated interesting ability and adaptability using an autoregressive approach, including few-shot capabilities, i.e. the ability to adapt to a new task with a limited number of examples (Brown et al. 2020). Though autoregressive LLMs are optimized for text generation by predicting the next token given some context, their classification performance has not yet been established.

While LLMs' word embeddings are difficult to interpret, the ability to use a prompt to direct the desired task and to inquire about the obtained output provides flexibility. However, recent findings support that foundation models can outperform domain-specific models in medical challenge benchmarks with appropriate prompting. Contrarily, BERT-based models are known to be computationally lightweight for fine-tuning and have easily accessible word embeddings for external analysis. On the other hand, autoregressive LLMs like GPT require substantial computational resources for both training and inference due to their significantly larger parameter count, which can be prohibitive for medical applications typically constrained to on-premise data storage and limited computing infrastructure.

Nonetheless, the application of all LLMs in the patient care domain is limited. One major obstacle is HIPAA regulation and privacy concerns since many models require uploading patient data to external host platforms (Z. Liu et al. 2023). Due to this limitation, many recent works have focused on locally fine-tuning BERT-like pre-trained models that work well on downstream tasks such as radiology report summarization ("Proceedings of the 20th Workshop on Biomedical Language Processing, BioNLP@NAACL-HLT 2021, Online, June 11, 2021" 2021), generation (Chen et al. 2020), and token-level or document-level classification (Jain et al. 2021; Chambon, Cook, and Langlotz 2023). However, these works typically do not provide diagnostic outputs, which are an important aspect of automated report classification. For instance, although rare attempts were made to classify lung diseases (n=14) using CheXbert, it only achieved a macro-averaged F1-score of 0.798, limiting its use in clinical settings (Smit et al. 2020).

The current work was done as part of the CARBI project investigating the causes of neurological complications in patients with acute respiratory failure (ARF) requiring invasive mechanical ventilation. ARF is the most common organ failure in the intensive care unit. It is associated with an increased risk of neurological complications (Battaglini D et al., 2020). When investigating causal relationships between potential insults that may have occurred during the patient's hospital stay and neurological complications, establishing the temporal relationship is crucial. However, most electronic health record-based datasets only provide diagnoses entered at the end of a hospital stay by medical archivists, without information about the precise date at which the diagnosis was made by the medical team, let alone potential classification errors. This can often complicate the use of these codes when exploring causal relationships. In contrast, radiology reports offer two distinct advantages: they are precisely dated and potentially more accurate, as medical imaging frequently serves as the primary basis for a wide array of diagnoses (Glance et al. 2006). In practice, when clinical suspicion of neurological complication arises, physicians generally perform brain imaging to assess structural damage.

To the best of our knowledge, no work has been done on classification of brain abnormalities in mechanically ventilated patients with ARF. Clinically relevant lesions that can appear on brain imaging studies (computed tomography and magnetic resonance imaging) performed in critically ill patients include cerebral infarcts, intraparenchymal, subarachnoid or intraventricular haemorrhage, cerebral edema, cerebral microbleeds, or anoxic-ischemic brain injury.

We therefore sought to develop a classification algorithm to automatically extract specific diagnoses from brain imaging reports, with the date of the report providing a fair estimation of the time at which the diagnosis was made.

**Methodology**

Data source:

We used the MIMIC–IV and MIMIC-IV-Note datasets for our analyses. These databases, hosted by the Laboratory for Computational Physiology at MIT, contain high-resolution information from hospital monitoring systems (including laboratory data, medication, and hospital administrative data), bedside monitoring systems (vital signs) along with certain caregiver notes and radiology reports from over 299,712 de–identified patients admitted to the Beth Israel Deaconess Medical Center between 2008 to 2019. It includes 2,321,355 time-stamped free-text reports of radiological exams (Johnson et al. 2023). For this work, we used all brain imaging studies performed in patients with ARF that underwent invasive mechanical ventilation and that did not have neurological injury upon ICU admission (n=219,532).

Initial Labelling

A sample of 1200 randomly selected reports were labelled independently by two human subject matter experts (PMW, NH, SADF) for the presence or absence of 11 non-mutually exclusive abnormalities: (1) Chronic Brain injury, (2) Acute Brain Injury (ABI), (3) Cerebral Infarct, (4) Subdural Haemorrhage, (5) Subarachnoid Haemorrhage (6) Intraparenchymal Haemorrhage, (7) Intraventricular Haemorrhage, (8) Anoxic-Hypoxic Brain Injury, (9) Brain Edema not otherwise specified, (10) Microbleeds, (11) Intracranial Hypertension. Conflicts were reviewed by senior author YAC.

Approach used: A first sample of 1000 reports was randomly divided into a training set (n=800), and validation set (n=200). We then proceeded to model development for our classification task in a step by step fashion described below, assessing model performance at each step.

## A0: RadBERT baseline results

In this work, we used RadBERT, a family of transformer-based language models adapted to radiology (Yan et al. 2022). RadBERT variants were pretrained with millions of radiology reports from the U.S. Department of Veterans Affairs (VA) healthcare system nationwide. We also provided a set of fine-tuning strategies that are helpful to optimize the performance of Neu-RadBERT for predicting the presence of brain abnormalities, thereby facilitating their classification outcomes. We used "out-of-the-box" RadBERT for the prediction of the aforementioned labels. We assessed the performance on the test dataset (number of reports=200).

## A1: Neu-RadBERT

This approach involved fine-tuning RadBERT on the training set and then using this trained model to analyze and predict the presence of brain abnormalities in the test and validation sets. The model architecture, hidden layers (12), attention heads (12), hidden size (768), and other hyperparameters were retained from the original model. We optimized for the number of training epochs (10), as further parameter tuning did not improve performance. The final model configuration remained largely consistent with the pre-trained version, ensuring compatibility with downstream applications.

## A2: Pretrained Neu-RadBERT with Masked Language Modelling

To improve performance, we pre-trained the model using Masked Language Modeling (MLM) on a dataset of approximately 2,000 unannotated radiology reports. In MLM, certain words in the text are concealed, and the model is trained to predict these concealed words. This process helps the model develop a robust understanding of medical terminology and the nuanced language associated with brain conditions, thus enhancing its predictive accuracy. After pre-training, the model underwent further training on a separate set of 800 annotated medical reports. This training also included a validation step on a subset of these reports, using an 80:20 split for training and validation. The fully trained model was then asked to make predictions on a validation set of 200 labelled reports, and the number of misclassified reports was recorded to assess performance.

<u>A3: Enhanced Pretrained Neu-RadBERT with Oversampling</u>

Building on A2, this approach addressed class imbalance — where certain labels like microbleeds were underrepresented in the dataset — through oversampling. For the oversampling of underrepresented classes in the training dataset, an additional 200 reports for these less common conditions were added to the training dataset (microbleed, subdural hematoma, anoxic brain injury, intracranial hypertension). These additional reports were found using a free search text strategy which contained the name of the brain abnormalities and their synonyms, which were again manually labelled. The oversampled dataset was also validated on a subset of these reports, using an 80:20 split for training and validation. This method aimed to improve Neu-RadBERT's ability to recognize and classify less common conditions more accurately, enhancing the model's overall diagnostic performance.

<u>B0: Llama-2-13B "zero-shot"</u>

This approach involved using "out-of-the-box" Llama-2 with 13B parameters for the binary prediction of acute brain injury versus no acute brain injury. The 13B model was chosen due to computational resource constraints. In one experiment, a zero-shot classifier pipeline was used to output a diagnostic label based only on an instruction and the input radiology report. The instruction provided each time was: "Assume you are a physician. I will transcribe a radiology report and you will tell me whether the report describes the presence of acute brain injury or no acute brain injury. Be concise: return only the label that best applies: 'acute' or 'not acute'."

<u>B0: Llama-2-13B "in-context"</u>

In a second experiment, a second classifier pipeline used the principle of few-shot learning using task specific prompts. It started with the instruction: "Consider the following two examples of reports and the expected label associated to each." A complete input-output example of an acute and chronic (not acute) brain injury radiology report followed. Then, the next instruction: "Now help me on this next radiology report with unknown label and predict the appropriate label." This was appended to the same instruction as in the first experiment before continuing with the input radiology report.

<u>B1: Llama-2-13B fine-tuned "zero-shot" and "in-context"</u>

This approach involved fine-tuning Llama-2-13B with our labelled report data using multiple hyperparameters to try to maximize classification performance. All newline characters were manually removed from the radiology reports. An appropriate cueing prompt was added to each diagnostic label in the output to provide context. Different training rank and learning rates were attempted. Zero-shot and in-context learning results were compared.

Computer resources

All experiments were performed on a A5000 NVIDIAgraphic processing unit with 24 Gb of VRAM, using python version 3.10. The analysis leveraged core libraries such as hugging face transformers for model fine-tuning and evaluation, PyTorch for deep learning computations, and hugging face datasets for dataset management. Additional libraries, including scikit-learn, pandas, and NumPy, were utilized for data preprocessing and evaluation metrics computation.

**Results**

Acute brain injury, particularly ischemic stroke and intraparenchymal hemorrhage were the most common abnormalities in the dataset (Table 1). In contrast, microbleeds were very rare before oversampling. This improved significantly after oversampling the reports containing this type of anomaly using a free-text search.

**Table 1. Frequency of positive cases in the test/internal validation set before after oversampling rare findings and in external validation dataset**

| Brain abnormality | Frequency of positive cases in labelled radiology reports | | |
| --- | --- | --- | --- |
| | Dataset before oversampling (random sample) | Dataset after oversampling | External validation set (random sample) |
| Chronic Brain Injury | 410/1000 (41%) | 496/1200 (41.3%) | 105/200 (52.5%) |
| Acute Brain Injury | 358/1000 (35.8%) | 491/1200 (40.9%) | 53/200 (26.5%) |
| Ischaemic stroke | 124/1000 (12.4%) | 157/1200 (13.1%) | 18/200 (9%) |
| Subdural hemorrhage | 29/1000 (2.9%) | 66/1200 (5.5%) | 4/200 (2%) |
| Subarachnoid hemorrhage | 67/1000 (6.7%) | 95/1200 (7.9%) | 11/200 (5.5%) |
| Intraparenchymal hemorrhage | 125/1000 (12.5%) | 163/1200 (13.6%) | 17/200 (8.5%) |
| Intraventricular hemorrhage | 79/1000 (7.9%) | 108/1200 (9%) | 18/200 (9%) |
| Anoxic brain injury | 50/1000 (5%) | 73/1200 (6.1%) | 5/200 (2.5%) |
| Brain edema | 67/1000 (6.7%) | 87/1200 (7.2%) | 5/200 (2.5%) |
| Microbleeds | 8/1000 (0.8%) | 55/1200 (4.6%) | 0/200 (0%) |
| Intracranial hypertension | 44/1000 (4.4%) | 48/1200 (4%) | 6/200 (3%) |

The implementation of the Neu-RadBERT model, refined through a series of strategic enhancements, yielded promising results in the classification of brain abnormalities on test data (Table 2). In cases of Acute Brain Injury, a dramatic improvement was observed after fine-tuning, with accuracy rising sharply from 45.1% to 99.0%, and the error rate dropping from 54.9% to 1.0%. Similar enhancements were seen across other brain abnormalities, with Ischemic stroke, Subdural Hematoma (SDH), Subarachnoid Hemorrhage (SAH), Intracerebral Hemorrhage (IPH), and Intraventricular Hemorrhage (IVH) all showing improvements after fine-tuning, presenting accuracies in the high 90s. Notably, classifications improved marginally for underrepresented brain abnormalities, namely Ischaemic stroke, SDH, SAH, IPH, Intracranial HTN following A3's oversampling strategy (Table 2).

**Table 2. Validation of the BERT-based models with the external dataset**

| Brain abnormality | Accuracy (% correct classifications) | | | |
|---|---|---|---|---|
| | **A0: Baseline** | **A1: Fine-Tuned** | **A2: Masked-Language** | **A3: Over-sampled rare events** |
| Chronic Brain Injury | 84.1% | 88.7% | **90.3%** | 89.2% |
| Acute Brain Injury | 45.1% | 99.0% | **99.5%** | 98.0% |
| Ischaemic stroke | 54.9% | 95.9% | 96.4% | **96.9%** |
| Subdural hemorrhage | 59.5% | 97.4% | 99.0% | **99.5%** |
| Subarachnoid hemorrhage | 54.4% | 98.0% | 97.4% | **99.0%** |
| Intraparenchymal hemorrhage | 56.4% | 94.4% | 96.4% | **97.4%** |
| Intraventricular hemorrhage | 54.9% | 95.4% | **97.4%** | **97.4%** |
| Anoxic brain injury | 57.4% | 97.4% | **98.5%** | **98.5%** |
| Brain edema | 58.5% | 97.4% | **97.9%** | 97.4% |
| Microbleeds | | | | |
| Intracranial hypertension | 57.2% | 96.4% | 97.4% | **98.0%** |

Before the implementation of oversampling, the dataset exhibited a significant imbalance with underrepresented positive brain abnormalities, this bias for the majority class inflated the F1-score in the validation process. We observed that this was particularly evident when the minority brain abnormality had very few positive instances; the model tended to ignore it, which did not substantially penalize the F1-score. Hence, complete (internal) performance of the A3 can be found in Table 3.

**Table 3: Complete Performance results for Model A3.**

| Brain abnormality | Strategy used | Epoch | Training Loss | Validation Loss | Accuracy | F1 Score | Precision | Recall |
|---|---|---|---|---|---|---|---|---|
| Chronic Brain Injury | A3 | 10 | 0.0087 | 0.3833 | 0.9625 | 0.9626 | 0.9637 | 0.9625 |
| Acute Brain Injury | A3 | 10 | 0.0174 | 0.0712 | 0.9917 | 0.9917 | 0.9916 | 0.9917 |
| Ischaemic stroke | A3 | 10 | 0.0386 | 0.0463 | 0.9917 | 0.9915 | 0.9917 | 0.9917 |

| Subdural hemorrhage | A3 | 10 | 0.2302 | 0.2652 | 0.9375 | 0.9073 | 0.8789 | 0.9375 |
| Subarachnoid hemorrhage | A3 | 10 | 0.0924 | 0.0897 | 0.9708 | 0.9723 | 0.9757 | 0.9708 |
| Intraparenchymal hemorrhage | A3 | 10 | 0.0202 | 0.1940 | 0.975 | 0.9747 | 0.9746 | 0.975 |
| Intraventricular hemorrhage | A3 | 10 | 0.0214 | 0.0750 | 0.9917 | 0.9915 | 0.9917 | 0.9917 |
| Anoxic brain injury | A3 | 10 | 0.0216 | 0.00007 | 1 | 1 | 1 | 1 |
| Brain edema | A3 | 10 | 0.2867 | 0.2856 | 0.9292 | 0.8950 | 0.8633 | 0.9292 |
| Microbleeds | A3 | 10 | 0.0505 | 0.07546 | 0.9875 | 0.9873 | 0.9873 | 0.9875 |
| Intracranial hypertension | A3 | 10 | 0.1615 | 0.0801 | 0.9708 | 0.9565 | 0.9425 | 0.9708 |

In each validation case, the F1- score must be considered in the context of the number of positive samples. Higher F1-scores in outcomes with fewer positive instances may imply that these conditions had distinctive features that the model could learn to recognize despite fewer training examples. However, it is also important to consider the potential for overfitting and to validate these findings with an independent test dataset, as shown in Table 1. For conditions with more positive instances, the model had more data available to learn from, which typically resulted in more reliable and generalizable learning, as reflected in the corresponding F1-scores.

As for Llama-2-13B, zero-shot binary classification accuracy was worse than guessing at 26.7% on the validation set. (Table 4) The addition of in-context learning increased classification accuracy to 34.17%. However, inspection of the predictions revealed collapse as all samples were classified as acute brain injury. Extensive fine-tuning and hyperparameter tuning on labelled data significantly improved classification accuracy to be somewhat better than guessing. There was no observed collapse with the in-context model. However, it is unlikely in-context learning leads to improvement for this specific task.

**Table 4**: **Binary classification accuracy for Llama-2-13B B0, B1**

| Brain abnormality | Accuracy (% correct classifications, % incorrect classifications) | | | |
| --- | --- | --- | --- | --- |
| | **B0 zero-shot** | **B0 in-context** | **B1 zero-shot** | **B1 in-context** |
| AcuteBI vs No AcuteBI | 26.7, 73.3 | 34.2, 65.8 | **67.5, 32.5** | 66.7, 33.3 |

**Discussion**

The direct application of Neu-RadBERT offers a quick, lightweight and straightforward method for disease classification, leveraging the model's inherent strengths. However, its effectiveness can be limited by the generic nature of its pretraining, which may not fully capture the specificities of medical language as related to neurological conditions. Enhancing the model with MLM pre-training on target domain tasks, boosted its medical report classification accuracy. Further augmenting the pretrained model with our oversampling technique addresses the challenge of class imbalance, ensuring that the model is well-equipped to recognize and accurately classify conditions that are underrepresented in the dataset. This approach tailored the model more closely to the domain-specific language and concepts found in brain imaging reports. Using this research tool, we were able to produce labels on the complete set of the reports for use in the CARBI study.

We think that this pre-trained model, made public on Hugging Face (https://huggingface.co/datasets/manisggn/Neu-Radbert/tree/main), can be utilised in various research contexts. By automatically producing categorical labels from radiology reports, such a strategy can be used to turn unstructured data into structured data that can be used in conventional statistical models. Moreover, such labels, with the level of accuracy achieved, could be in turn used to train computer vision classification algorithms. This is crucial for a comprehensive and balanced diagnostic tool that can reliably support medical professionals across a spectrum of brain-related conditions.

Llama-2

We hypothesize that the changes introduced during in-context learning and model fine-tuning confuse Llama-2 due to insufficient training data for the complexity of the classification task. In addition, the model's understanding of medical vocabulary during its base training may also have been inadequate. Further foundational training of Llama-2 on medical corpora could be helpful before task-specific training. The poor results obtained discouraged us from pursuing further experimentation with autoregressive LLMs. It is possible that larger foundation models would offer improved performances at a higher computational cost. Potential future work could include the comparison of different parameter-efficient fine-tuning strategies to decrease the domain adaptation computational cost.

**Limitations**

External validation on a separate dataset was not performed, which could further help assess the generalizability of our approach. Additionally, alternative open-source LLMs were not explored, leaving room for future comparisons. The impact of computational constraints on model performance was not systematically evaluated and could be an area for further optimization.

**Conclusion**

For the specific task of classifying radiology reports for brain injuries in the setting of ARF, we found that a specialized BERT-based model could achieve great accuracy that we could not match with a newer general-purpose autoregressive LLM. We used different methodologies to address challenges in the diagnostic process. By tailoring the model to better understand neuroradiological language and by ensuring a balanced approach to data representation, we were able to significantly enhance the accuracy and reliability of the NLP model to automate report-level classification labelling. Neu-RadBERT's application in diagnosing brain injury demonstrates the potential of NLP technologies to revolutionize medical research by allowing to transform widely available, but mainly unused, unstructured data residing in the free text form, into structured data that can readily be used. Future research and development should focus on further optimising these methodologies, potentially incorporating more advanced techniques and broader datasets.

## Appendix

## B0 base Llama-2-13B in-context sample example:

"instruction": "Assume you are a physician. I will transcribe a radiology report and you will tell me whether the report describes the presence of acute brain injury or no acute brain injury. Be concise: return only the label that best applies: 'acute' or 'not acute'."

Consider the following two examples of reports and the expected label associated to each.

"input": "HISTORY: ___ female with left posterior communicating artery aneurysm with worsening neurologic examination. Evaluation for interval change. TECHNIQUE: Contiguous axial images were obtained through the brain without the administration of IV contrast using a portable CT scanner. COMPARISON: Comparison is made to CTA of the head and neck from ___. FINDINGS: Compared with the prior study, there has been some redistribution of previously identified subarachnoid hemorrhage in the left sylvian fissure, obscuring the sulci. Subarachnoid hemorrhage is also re-demonstrated within the interpeduncular cistern, ambient cisterns, and is again seen layering in the occipital horns of both lateral ventricles, which is more apparent than on the prior study. There is no evidence of midline shift. No new areas of hemorrhage are identified. There is no evidence of edema, mass, mass effect or infarction. Nasal and endotracheal tubes are in place. The visualized paranasal sinuses, mastoid air cells and middle ear cavities are clear. No cranial or facial soft tissue abnormalities are present. IMPRESSION: 1. Interval redistribution of previously identified subarachnoid hemorrhage within the left sylvian fissure, basal cisterns, and layering in the lateral ventricles posteriorly. 2. No new areas of hemorrhage are identified."
"output": "The type of brain injury is: acute"

"input": "HISTORY: Piriform sinus injury and retroesophageal abscess, status post incision and drainage. Evaluate for abscess. TECHNIQUE: Contiguous axial MDCT images were obtained through the brain without administration of IV contrast. Reformatted coronal and sagittal and thin section bone algorithm reconstructed images were acquired. DLP: 1003.42 mGy-cm. COMPARISON: Outside hospital CT neck from ___ on ___. FINDINGS: There is no acute hemorrhage, edema, mass effect, or large territorial infarct. The ventricles and sulci are prominent, suggestive of age-related volume loss. Periventricular white matter hypodensities are consistent with chronic small vessel ischemic disease. Mucosal thickening and fluid levels within the visualized paranasal sinuses are likely secondary to intubation. The mastoid air cells and middle ear cavities are clear. Air within the soft tissues is better evaluated on CT neck from the same day. IMPRESSION: No acute intracranial abnormality. Air within the soft tissues is better evaluated on CT neck from the same day."
"output": "The type of brain injury is: not acute"

Now help me on this next radiology report with unknown label and predict the appropriate label.

**[INSERT RADIOLOGY REPORT TO CLASSIFY]**

Four representative sample outputs with associated probabilities (reports redacted):

\'output\': \'The type of brain injury is: \'}]', 'labels': ['acute', 'acute'], 'scores': [0.7102524757385254, 0.289747554063797]}

\'output\': \'The type of brain injury is: \'}]', 'labels': ['acute', 'not acute'], 'scores': [0.6549529194831848, 0.3450471103191376]}

\'output\': \'The type of brain injury is: \'}]', 'labels': ['acute', 'acute'], 'scores': [0.6401861310005188, 0.3598138988018036]}

\'output\': \'The type of brain injury is: \'}]', 'labels': ['acute', 'not acute'], 'scores': [0.7309661626815796, 0.2690338671207428]}

The model is somewhat falsely confident in predicting samples as all examples of acute brain injury. The analogous zero-shot B0 model presented probability scores much closer to 50%, indicating a lack of confidence in its predictions.

B1 fine-tuned Llama-2-13B in-context sample example (rank 16, 2e-4 learning rate, 1920 epochs of n=786, 4.5 hours running time):

**[Same instruction as analogous B0 configuration]**

Four representative sample outputs with associated probabilities (reports redacted):

\'output\': \'The type of brain injury is: \'}]', 'labels': ['not acute', 'acute'], 'scores': [0.5146442651748657, 0.4853557348251343]}

\'output\': \'The type of brain injury is: \'}]', 'labels': ['acute', 'not acute'], 'scores': [0.5370413661003113, 0.46295860409736633]}

\'output\': \'The type of brain injury is: \'}]', 'labels': ['acute', 'not acute'], 'scores': [0.581303060054779, 0.41869688034057617]}

\'output\': \'The type of brain injury is: \'}]', 'labels': ['not acute', 'acute'], 'scores': [0.5273165106773376, 0.47268351912498474]}

The model is less confident with its choice after fine-tuning than before, with output probabilities hovering much closer to 50% for both the zero-shot and in-context models.

Exploration of data augmentation using base Llama-2-13B to generate 800-token report examples of under-represented diagnoses in the training set (epidural and subdural hemorrhage, microbleeds, anoxic brain injury, intracranial hypertension): Instruction: "Assume you are a physician. Create a realistic and complete radiology report that describes **[INSERT DIAGNOSIS HERE]**. Make sure to include appropriate indication, findings and impression sections in your report."

<s> Assume you are a physician. Create a realistic and complete radiology report that describes epidural hemorrhage. Make sure to include appropriate indication, findings and impression sections in your report.

Radiology Report:

Patient Name: John Doe

Date of Exam: 2/12/2023

Indication:

The patient presented with severe headache, nausea, and vomiting after a fall. The patient's vital signs were: Blood pressure 180/100 mmHg, Pulse 120/min, Respiratory rate 24/min, Temperature 37.2°C (99°F).

Findings:

1. Epidural hemorrhage: A large epidural hemorrhage is present in the left posterior fossa, measuring approximately 3 cm in diameter. The hemorrhage is isodense on CT scan and does not demonstrate any areas of calcification or bone erosion.
2. Skull fracture: A linear skull fracture is present in the left parietal region, measuring approximately 2 cm in length. The fracture is displaced and is associated with a small amount of subdural hematoma.
3. Intracerebral hemorrhage: A small area of intracerebral hemorrhage is present in the left frontal lobe, measuring approximately 1 cm in diameter.
4. Subdural hematoma: A small area of subdural hematoma is present in the left frontal lobe, measuring approximately 1 cm in diameter.
5. No evidence of intracranial air or blood.

Impression:

1. Epidural hemorrhage with skull fracture and intracerebral hemorrhage.
2. Acute subdural hematoma.
3. No evidence of intracranial air or blood.

Recommendations:

1. Immediate surgical intervention is recommended to evacuate the epidural hemorrhage and relieve pressure on the brain.
2. Close monitoring of the patient's neurological status and vital signs is recommended.
3. CT scans of the brain should be performed serially to monitor for any changes in the hemorrhage and to assess for any new areas of bleeding

<s> Assume you are a physician. Create a realistic and complete radiology report that describes subdural hemorrhage. Make sure to include appropriate indication, findings and impression sections in your report.

Radiology Report:

Patient Name: John Doe

Date of Exam: 2/12/2023

Indication: Headache, vomiting, and decreased level of consciousness

Findings:

1. Subdural hemorrhage: A well-defined, hyperdense lesion is observed in the left posterior frontal lobe, measuring approximately 2 cm in diameter. The hemorrhage is isointense to blood on T2-weighted imaging and hyperintense on T1-weighted imaging, consistent with recent bleeding. There is surrounding edema and vasogenic effusion in the left posterior frontal lobe.

2. Mass effect: The subdural hemorrhage is causing significant mass effect on the surrounding brain tissue, particularly in the left posterior frontal lobe, resulting in compression and displacement of the brain tissue.

3. Cerebral edema: There is evidence of cerebral edema in the left posterior frontal lobe, as indicated by the hyperintense signal on T2-weighted imaging.

4. Vasogenic effusion: There is evidence of vasogenic effusion in the left posterior frontal lobe, as indicated by the presence of extracellular fluid on T2-weighted imaging.

Impression:

1. Subdural hemorrhage: A well-defined, hyperdense lesion is present in the left posterior frontal lobe, consistent with recent bleeding.

2. Mass effect: The subdural hemorrhage is causing significant mass effect on the surrounding brain tissue, resulting in compression and displacement of the brain tissue.

3. Cerebral edema: There is evidence of cerebral edema in the left posterior frontal lobe.

4. Vasogenic effusion: There is evidence of vasogenic effusion in the left posterior frontal lobe.

Recommendations:

1. Urgent neurosurgical intervention is recommended to evacuate the subdural hemorrhage and relieve the mass effect on the surrounding brain tissue.

2. Close monitoring of the patient's clin

<s> Assume you are a physician. Create a realistic and complete radiology report that describes microbleeds. Make sure to include appropriate indication, findings and impression sections in your report.

Microbleeds are small hemorrhages that occur in the brain and are often seen on magnetic resonance imaging (MRI) scans. They can be caused by a variety of factors, including trauma, hypertension, and certain medical conditions such as sickle cell disease. Microbleeds can be difficult to detect and diagnose, and they can have significant implications for patient outcomes.

Here is a sample radiology report that describes microbleeds:

Indication:
The patient presented with a headache, dizziness, and confusion. The patient also reported a recent fall and hit their head.

Findings:
MRI brain with contrast was ordered to evaluate for intracranial hemorrhage. The MRI scan revealed multiple small hemorrhages in the bilateral cerebral hemispheres, predominantly in the frontal and parietal lobes. These hemorrhages are consistent with microbleeds. There is no evidence of acute intracranial hemorrhage or mass effect.

Impression:
The findings of multiple microbleeds in the bilateral cerebral hemispheres are consistent with a history of trauma and hypertension. These microbleeds can cause significant damage to the brain over time and may lead to long-term cognitive and functional decline. It is important for the patient to be closely monitored and managed appropriately to prevent further complications.

Recommendations:
The patient should be referred to a neurologist for further evaluation and management. Regular follow-up MRI scans should be performed to monitor for any changes in the microbleeds. The patient should be advised to avoid any activities that may exacerbate the condition, such as heavy lifting or bending.

Key Points:

* Microbleeds are small hemorrhages that can occur in the brain and can be caused by a variety of factors.
* Microbleeds can be difficult to detect and diagnose, and they can have significant implications for patient outcomes.
* MRI scans can be used to evaluate for microbleeds and monitor for any changes over time.
* Patients with microbleeds should be closely monitored and managed appropriately to prevent further complications

<s> Assume you are a physician. Create a realistic and complete radiology report that describes anoxic brain injury. Make sure to include appropriate indication, findings and impression sections in your report.

Anoxic Brain Injury

Patient Name: [Insert patient name]

Date of Exam: [Insert date of exam]

Indication:

The patient presented with a history of cardiac arrest and anoxia for [insert time period] minutes. The patient was resuscitated and brought to the emergency department with a Glasgow Coma Scale (GCS) of [insert GCS score]. The patient's family reported that the patient had been in a coma for the past [insert time period] days.

Findings:

1. Non-contrast CT scan of the brain was performed to evaluate for anoxic brain injury.

2. The scan revealed a large area of infarction in the left hemisphere, predominantly involving the anterior cerebral cortex and subcortical regions. The infarction is consistent with anoxic injury due to lack of blood flow.

3. There is evidence of cytotoxic edema in the affected regions, with corresponding increased intensity on FLAIR and T2-weighted sequences.

4. No evidence of intracerebral hemorrhage or mass effect was observed.

Impression:

1. Anoxic brain injury with left hemispheric infarction and cytotoxic edema.

2. The patient's GCS score of [insert GCS score] and prolonged anoxia suggest a severe injury with potential for long-term neurological sequelae.

3. Further imaging and clinical follow-up are necessary to assess for any further changes in the patient's neurological status and to determine the appropriate course of treatment.

Recommendations:

1. The patient should be closely monitored for any changes in neurological status, including fluctuations in GCS score and evidence of new infarction or edema.

2. The patient should be treated with appropriate supportive care, including mechanical ventilation and vasopressors as needed.

3. The patient should be considered for therapeutic hypothermia to reduce the risk of secondary injury and improve outcomes.

4. Further imaging, such as MRI or diffusion-

<s> Assume you are a physician. Create a realistic and complete radiology report that describes intracranial hypertension. Make sure to include appropriate indication, findings and impression sections in your report.

Radiology Report:

Patient Name: John Doe

Date of Exam: 2/12/2023

Indication:

The patient presented with a headache, nausea, and vomiting. The patient's physician suspected intracranial hypertension and ordered a CT scan of the brain.

Findings:

1. The patient's non-contrast CT scan of the brain revealed no acute intracranial abnormalities, such as hemorrhage or mass lesions.
2. There was evidence of intracranial hypertension, including dilated cerebral vasculature and prominent sulci.
3. The patient's ventricular system was enlarged, with a prominent ventricular system and dilated cerebral aqueduct.
4. There was no evidence of hydrocephalus or other intracranial abnormalities.

Impression:

Based on the patient's symptoms and the findings of the non-contrast CT scan of the brain, it is impression that the patient has intracranial hypertension. The dilated cerebral vasculature and prominent sulci are consistent with this diagnosis. The patient's ventricular system is enlarged, which is also consistent with intracranial hypertension. However, there is no evidence of hydrocephalus or other intracranial abnormalities.

Recommendations:

1. The patient should be referred to a neurologist for further evaluation and management of intracranial hypertension.
2. The patient should be monitored closely for any changes in symptoms or signs of intracranial hypertension.
3. The patient should be considered for imaging studies, such as MRI or MRA, to further evaluate the cause of intracranial hypertension.

Keywords: intracranial hypertension, dilated cerebral vasculature, prominent sulci, enlarged ventricular system, hydrocephalus.</s>